\documentclass[twocolumn,preprintnumbers,aps,prl,amsmath,amssymb]{revtex4}
\usepackage{graphicx}
\usepackage{dcolumn}
\usepackage{bm}

\begin{document}

\newcommand{\vo}{VO$_x$}
\title{Giant positive magnetoresistance in metallic VO$_{x}$ thin films}

\author{A. D. Rata$^1$, V. Kataev$^{2,3}$, D. Khomskii$^{2,4}$, and T. Hibma$^1$}
\address{$^1$Chemical Physics Department, Groningen University, Nijenborgh 4, 9747 AG Groningen, The Netherlands}
\address{$^2$\textrm{II}. Physikalisches Institut, Universit\"{a}t zu K\"{o}ln, Z\"{u}lpicher Str. 77, 50937 K\"{o}ln, Germany}
\address{$^{3}$Kazan Physical Technical Institute, Russian Academy of Sciences, 420111 Kazan, Russia}
\address{$^{4}$ Laboratory of Solid State Physics, Groningen University, Nijenborgh 4, 9747 AG Groningen, The Netherlands}

\date{\today}

\begin{abstract}
We report on giant positive magnetoresistance effect observed in
VO$_{x}$ thin films, epitaxially grown on SrTiO$_{3}$ substrate.
The MR effect depends strongly on temperature and oxygen content
and is anisotropic. At low temperatures its magnitude reaches
$70\%$ in a magnetic field of $5$ T. Strong electron-electron
interactions in the presence of strong disorder may qualitatively
explain the results. An alternative explanation, related to a
possible magnetic instability, is also discussed.
\end{abstract}
\maketitle

Transition metal oxides (TMO) with strong electronic correlations
show many fascinating phenomena, like high temperature
superconductivity, spin-charge separation,  quantum criticality,
or colossal negative magnetoresistance. The rich physics of TMO is
due to strong coupling of  spin, orbital and lattice degrees of
freedom. Their complex interplay is controlled by a large number
of structural and chemical factors, which makes the search for the
new oxide materials, where the fine tuning of chemical composition
and steric parameters might yield new unexpected electronic and
magnetic properties, of an apparent fundamental and applied
interest.

A quite unexplored class of oxides containing transition metal
ions are the monoxides of vanadium and titanium. In these oxides
the overlap of the metal $t_{2g}$ orbitals can be tuned by
changing the oxygen stoichiometry, which offers the control of the
width of the $t_{2g}$ band responsible for electron conductivity.
Owing to a broad characteristic stoichiometry range, small changes
in composition and/or local geometry can induce rather diverse
physical properties. The oxygen content $x$ in vanadium monoxide
VO$_{x}$, may deviate substantially from $1$
($0.8\!<\!x\!<\!1.3$). Remarkably, already small variations of $x$
can induce changes in the electrical conductivity: A gradual
transition from a metallic to a semiconducting behavior has been
observed in the bulk \cite{banus} and in thin films \cite{rata}.
Another important feature is an intrinsic disorder. Even
stoichiometric VO contains $\sim 16\%$ atomic vacancies in both
sublattices, distributed at random.

In this paper we report on giant {\em positive}
magnetoresistance (up to $70\%$ in a magnetic field of $5$~T)
observed at low temperatures in compressively strained metallic
VO$_x$ thin films ($0.8\!<\!x\!<\!1$). The positive sign of the
effect, i.e. the {\em increase} of the resistance with applied
magnetic field, implies different underlying physics as
compared to metallic multilayers or manganite perovskites,
where, as a rule, the resistance {\em decreases} with field
\cite{note}. We discuss possible mechanisms of this unusual
effect, which might be related to electron-electron interaction
in the presence of strong disorder, or to the proximity of the
system to magnetic instability.

The $100~\AA$ thick VO$_x$ films were grown on SrTiO$_{3}$
(001)(STO) substrates by Molecular Beam Epitaxy. To prevent
after-oxidation, the films were capped with a thin MgO layer.
The oxygen content $x$ was determined using
$^{18}$O$_{2}$-Rutherford Backscattering Spectrometry, as
described in Ref.~2. Single-phase material was obtained for
$x$-values varying from $0.8$ to $1.22$. \textit{In-situ} RHEED
proved the \textit{layer-by-layer} growth, with the same
orientation as the underlying substrate. The high quality of
our samples was further confirmed by x-ray analysis, which
shows that the film grows in a full coherence with the
substrate, i.e the in-plane film and substrate lattice
constants are identical ($3.903~\AA$). The out-of-plane lattice
constant varies between $4.003~\AA$ for $x\!=\!0.8$ and
$3.974~\AA$ for $x\!=\!1.22$. Resistance and magnetoresistance
(MR) were measured by the standard four-point probe method, in
a commercial PPMS system, equipped with a rotatable sample
holder, at temperatures between 2 and 300~K. For the MR
measurements the magnetic field was varied between 5~T and
-5~T, applied either perpendicular to the film plane, or in the
film plane and parallel to the current. The Hall coefficient
was measured in the square geometry using a conventional $ac$
bridge. Electrical contacts of 10~nm of Cr metal were
evaporated on the STO substrates, prior to deposition of the
VO$_x$ layers.

In a previous study we found that the resistivity of VO$_x$ films
grown under tensile strain on MgO substrates is orders of
magnitude larger than for bulk material of the same composition
\cite{rata}. Consistently, in this study we find much lower
resistivities for VO$_x$ films on STO. Presumably, this is due to
the increase in the direct overlap between t$_{2g}$ orbitals of
neighboring metal ions in the compressively strained films. From
Fig.~1 it is evident that there is a gradual transition from
metallic to semi-conducting behavior at about $x = 0.94$.

In Fig.~2 we compare the temperature dependence of the resistivity
$\rho$ of the VO$_{x}$ sample with $x\! =\! 0.82$ measured in zero
magnetic field $H$ with that in an applied field of $5$~T.
$\rho(T)$ strongly decreases with decreasing temperature, reaches
a minimum with the value of $\sim8 \mu \Omega cm$ at $T\! = 25$ K,
and increases steeply at still lower temperatures.  The strong
$T$-dependence of $\rho$ and extremely low residual resistivity
implies that the electronic conduction is mainly due to
electron-electron scattering. The decrease of conductivity related
to the steep upturn of $\rho(T)$ below $25$ K follows very well a
$\sqrt{T}$ law (see left inset in Fig.~2). As to the field
dependence, the two curves, $\rho(T,H\!=\!0)$ and $\rho(T,H\!=\!5$
T) are indistinguishable down to $\sim\! 70$ K. Surprisingly, at
lower temperatures the application of magnetic field {\em
increases} the resistivity.

\begin{figure}
\includegraphics[angle=-90,width=0.8\columnwidth]{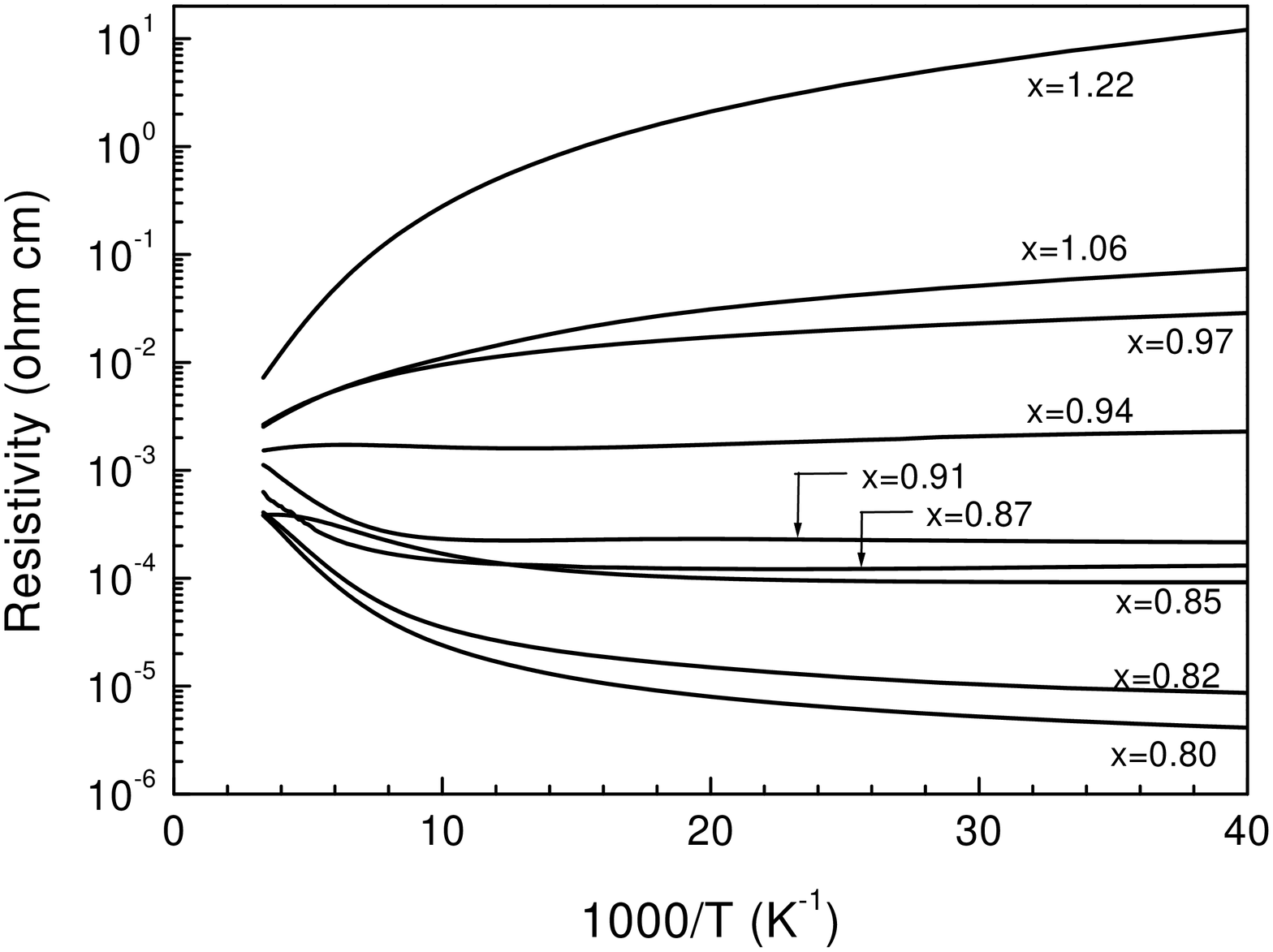}
\caption{ Resistivity of VO$_x$/STO films for different x. }
\label{res}
\end{figure}

In the inset at the right of Fig.~2 we show the temperature
dependence of MR defined as
$\{[\rho(H)-\rho(0)]/\rho(0)\}\!\times\! 100\%$. The sign of the
MR is positive and its magnitude increases smoothly with
decreasing temperature, both in a drastic contrast to the negative
colossal MR observed in the manganites \cite{schiffer}.

\begin{figure}[h]
\includegraphics[angle=-90,width=0.8\columnwidth]{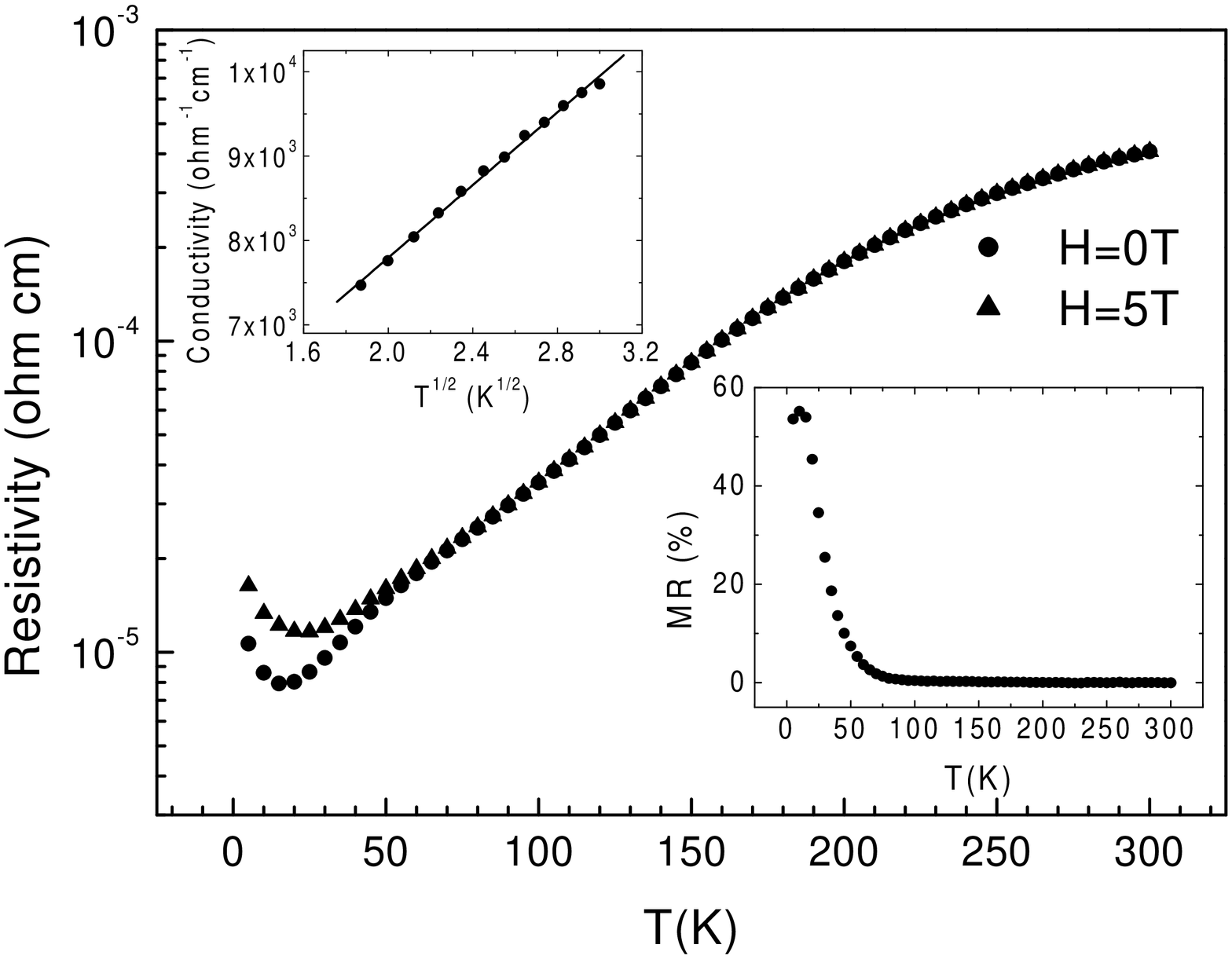}
\caption{The resistivity of VO$_x$ film with x =0.82 at zero field
and at H = 5~T. In the inset at the right, the magnetoresistance
is shown. In the inset at the left we show the fit of the
low-temperature increase of resistivity by the $1/\sqrt{T}$ law.}
\label{fieldres}
\end{figure}

 A representative field dependence of MR for the $x\!=\! 0.85$ sample
at several selected temperatures is shown in Fig.~3a. The curves
were obtained in all cases by increasing the field from zero to
5~T, sweeping it then to -5~T and turning $H$ finally back to
zero. Under field cycles, the MR shows no hysteresis. Moreover, no
sign of saturation is observed up to 5~T. Except for small fields,
the MR is proportional to $\sqrt{H}$ at low temperatures and
gradually changes to an almost linear and finally quadratic
dependence with increasing temperature (see Fig.~3c).

A remarkable feature of the positive MR in VO$_x$ films is its
strong dependence on the oxygen stoichiometry (Fig.~3b). The
effect is largest in the sample with the smallest oxygen content
$x\!=\! 0.8$, where the MR amounts to 70~\% at 5~K in a magnetic
field of 5~T. Increase of $x$ results in the decrease of the MR
which finally becomes unobservable for $x\!>\! 0.94$, in an
apparent correlation with the crossover from metallic to
semiconducting regime of conductivity.

\begin{figure}
\includegraphics[angle=-90,width=\columnwidth]{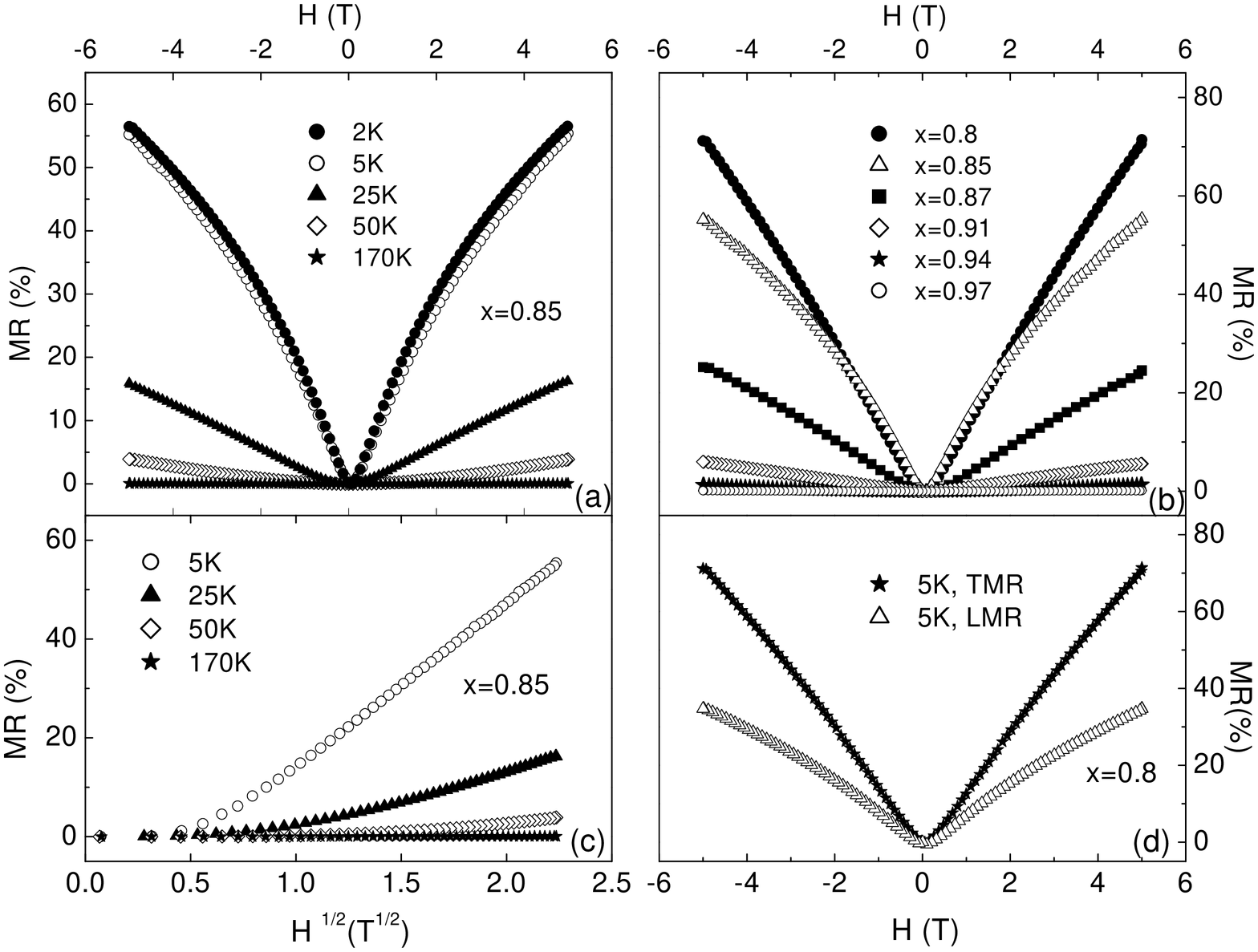}
\caption{(a) The magnetoresistance of VO$_x$ film with $x = 0.85$
for different temperatures. (b) The magnetoresistance of VO$_x$
films at 5 K for different $x$. (c) $\sqrt{H}$ dependence of MR at
low temperatures. (d) Comparison of the transverse and
longitudinal MR for $x = 0.8$ at 5 K.} \label{res}
\end{figure}

A clear dependence of the MR on the direction of the applied field
was found. In Fig.~3d we present the MR of the $x\! =\! 0.8$
sample at $T\!=\!5$~K with  H perpendicular and parallel to the
film plane and current direction. Positive transverse
magnetoresistance (TMR) of about 70~\% and a longitudinal, still
positive, magnetoresistance (LMR) of 40~\% was observed in a
magnetic field of up to 5~T. Note that the TMR is always larger
than the LMR.

The behavior of the MR is clearly correlated with the Hall effect.
Our measurements show that the Hall effect is quite strong,
negative and linear in fields up to $5$~T. In Fig.~4 we plot the
carrier concentration and Hall mobility obtained from the Hall
effect and resistivity data of the sample with $x\!=\!0.87$. One
can see that in the region of the strong change of the MR (for $T
< \sim 50 K$) the carrier concentration is essentially constant,
whereas the mobility which at low temperatures is pretty high,
starts to decrease.

\begin{figure}
\includegraphics[angle=-90,width=0.8\columnwidth]{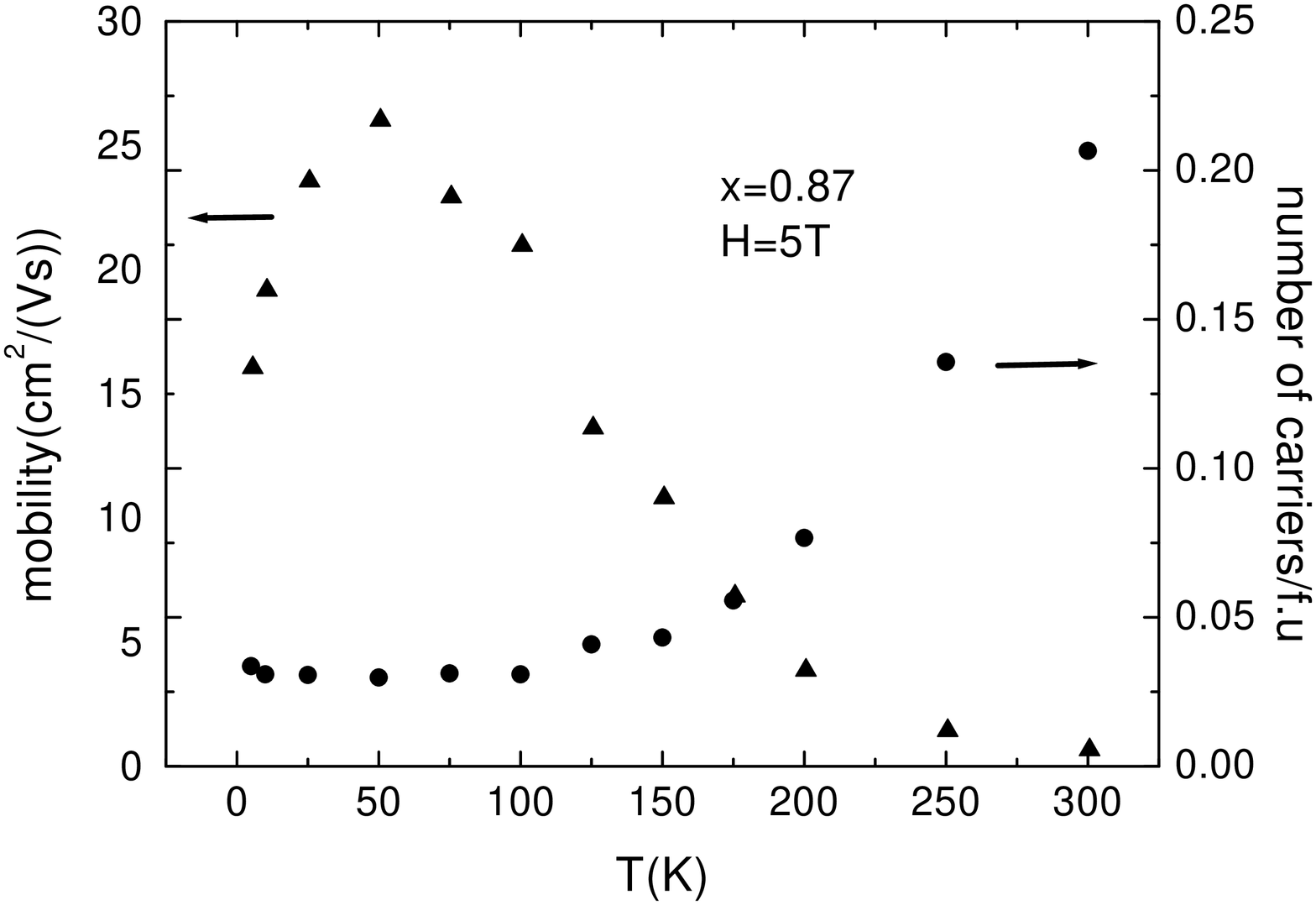}
\caption{The number of carriers/f.u and the Hall mobility of
VO$_x$ film with x =0.87 as the function of temperature in the
field of 5~T.} \label{fieldres}
\end{figure}

To elucidate magnetic properties of the VO$_x$ films we have
measured electron spin resonance (ESR) of the samples with $x$
ranging from 0.8 to 1.2. ESR has been measured using a Bruker
spectrometer at X-band frequency 9.48 GHz and at temperatures
between 1.9 and 300~K. A set of representative spectra (field
derivatives of the absorbed microwave power dP(H)/dH) is shown in
Fig.~5. The fingerprint of the VO$_x$/STO samples with $x<1$
showing a large positive MR is the occurrence at low temperatures
of an intense broad microwave absorption consisting of several
overlapping peaks. The signal emerges below 20-25 K. At still
lower temperatures it increases in intensity and acquires a
structure (Fig.~5a). Remarkably, in the same temperature interval
the resistivity shows an upturn (Fig.~2). In contrast, the samples
with $x>1$ as well as the STO substrate itself are ESR silent and
show only a small spurious signal at $H_{res}\simeq$ 3.3 kOe
(Fig.~5b), which can be attributed to the small amount of
paramagnetic impurities in the substrate.

The signal observed in the VO$_x$ films ($x<1$) is reminiscent of
ferromagnetic resonance in strongly inhomogeneous ferromagnetic
films, like, e.g., as-grown manganite films \cite{Lofland}. Owing
to an inhomogeneous distribution of the magnetization across the
film multiple broad lines occur in the spectrum. Apparently, a
similar spectrum might be expected if the sample is not yet
ferromagnetic but contains randomly distributed mesoscopic spin
clusters.

The positive magnetoresistance which we found in the VO$_x$ films
is quite unusual in several respects. First of all,  surprising is
the very large magnitude of the effect. The ordinary positive MR
in metals is usually rather small, less than a few percent. Its
size is determined by $\omega_{c}$$\tau$, where
$\omega_{c}=eH/m^{*}c$ is a cyclotron frequency, and $\tau$ the
relaxation time, which is proportional to the mean free path $l$.
Here $e$ is the electron charge, m$^{*}$ is the effective carrier
mass and $c$ is the speed of light. One may conclude that in our
sample the value of $l$ is rather large, as the residual
resistivity is quite low. Still, it is difficult to expect a MR of
$70\%$. Besides, the MR is also very large ($\sim 40\%$) for the
parallel field geometry ( see Fig.~3d), for which one would not
expect a strong orbital contribution in our thin films.

Another feature which is different in our samples is the field
dependence. The conventional MR in metals is quadratic in field
(only at the ultraquantum regime it may be linear
\cite{abrikosov}), whereas as one sees from Fig.~3, it is not the
case in our VO$_x$ films. As is discussed above, the character of
the field dependence changes gradually with increasing $T$ from
square root, to linear and finally to quadratic. In particular,
the linear regime resembles the MR in nonstoichiometric
Ag$_{2\pm\delta}$Se, Ag$_{2\pm\delta}$Te \cite{xu} (for which
positive MR is linear in field and larger in magnitude).

\begin{figure}[t]
\includegraphics[angle=-90,width=0.8\columnwidth]{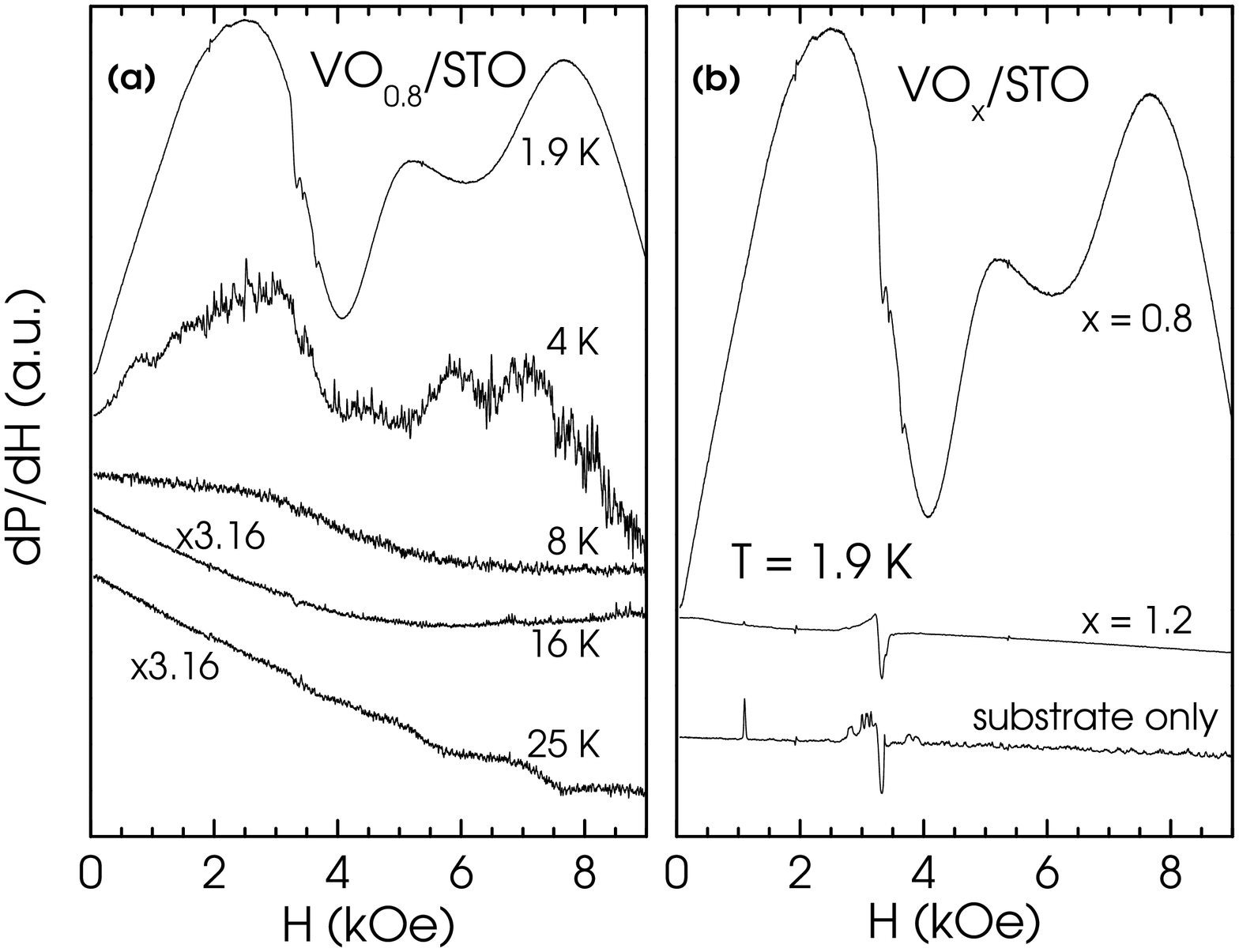}
\caption{ESR spectra of VO$_x$/STO samples. (see text)}
\label{ESR}
\end{figure}

The explanation of the unusual behavior of the MR in VO$_x$ films
may be thought in the specific electronic and crystal structure of
this compound. On the one hand, as always in transition metal
oxides, the electrons are apparently rather strongly correlated.
On the other hand, there exists strong intrinsic disorder in
VO$_x$, which always contains about $10 - 20\%$ of vacancies in
both anion and cation sublattices. The role of the structural
disorder has been recently discussed by Goodenough {\it et al.}
\cite{goodenough}. Thus, altogether one should view this system as
the one with strong disorder and strong interaction.

Theory predicts (see e.g. Ref.~\onlinecite{altshuler}) that in
metals with the electron-electron interaction and disorder, a
square-root singularity appears  in the density of states at
the Fermi level, with the corresponding anomalies in transport
properties, including MR. The correlation between the behavior
of the MR in VO$_x$ films with that of the Hall effect,
discussed above, suggests that indeed the interplay between
interaction and disorder may be decisive for the observed
effects. Thus, the low-temperature resistivity of our metallic
films of VO$_x$ is $\sim 1/\sqrt{T}$ (see left inset in
Fig.~2), which agrees with this model. The MR should behave as
$\sqrt{H}$ in high fields, at least for the case of relatively
weak disorder and interaction, treated in these papers
\cite{altshuler}. Our MR effect is qualitatively similar to
this behavior, although it is much stronger than those expected
from theoretical considerations \cite {*}. One reason for this
may be the much stronger disorder and interaction in our
system. The effect may be also enhanced by the paramagnon
scattering \cite{Kastrinakis}.

In spite of the qualitative similarity with the case of the weakly
interacting disordered systems, one still can not exclude
alternative explanations of the observed large positive MR in the
thin films of VO$_x$. In particular, one can expect that our
system may be rather close to magnetic instability. Relatively
broad bands and/or strong disorder may prevent the formation of
long-range magnetic ordering, although short-range magnetic
correlations may still exist. Indeed, our ESR study of VO$_x$
films has shown that a rather strong but quite unusual (broad,
consisting of several overlapping lines) ESR signal appears, which
one might expect from, e.g., random magnetic clusters.

Our situation is definitely different from, e.g., that in
phase-separated manganites (see, e.g.,
Ref.~\onlinecite{dagotto,khomskii1}), where the presence of
"preformed" ferromagnetic metallic clusters and their growth in
size with field leads to the colossal {\em negative}
magnetoresistance. However, we can visualize the following
scenario: In our inhomogeneous system there may be no preformed
magnetic clusters, but the magnetic susceptibility $\chi$ may be
strongly spatially  inhomogeneous, owing to, e.g., strong
disorder. Then in the external field the parts of the film with
larger $\chi$ will develop larger magnetization. But according to
the conventional double-exchange model the energy of conduction
electrons in these magnetized regions would decrease, and the
electrons of our sample would redistribute: The electron
concentration in these regions would increase, whereas the regions
in between would be depleted. If the system is still below a
percolation threshold, this would lead to the total {\em increase}
of resistivity, owing to creation of more insulating barriers
\cite{khomskii2}, i.e. to the total {\em positive} MR. This
picture is also consistent with the ESR data, although it is not
clear whether it would give the observed dependence of the MR on
$H$ and $T$.

To conclude, we observed a surprisingly large (up to 70\% in the
field of 5~T) positive magnetoresistance in thin VO$_x$ films,
grown epitaxially on the SrTiO$_3$ substrate, which behaves at low
temperatures as $\sqrt{H}$ in high fields. We argue that the
possible explanation of the observed behavior may rely on the
interplay of the electron-electron interaction and disorder. The
effect observed above is much stronger than that predicted
theoretically, but this can in principle be connected with the
much stronger interaction and disorder in our case. Still,
alternative explanations, e.g. relying on the inhomogeneous
magnetic susceptibility and the proximity of our system to
magnetic ordering, can not be excluded. In summary, the
compressively strained VO$_x$ system seems to be quite unusual in
many respects, and its properties, especially an extremely strong
positive magnetoresistance, deserve further study, and possibly
can be useful in applications, e.g. in spintronics.

We are grateful to T. T. M. Palstra, R. de Groot, L. H. Tjeng, P.
A. Lee, I. Elfimov, G. A. Sawatzky, P. Littlewood, and M. V.
Mostovoy for useful discussions. This work was supported by  the
Deutsche Forschungsgemeinschaft through SFB 608 and by the
Netherlands Foundation for Fundamental Study of Matter (FOM).

\end{document}